# Observation of second sound in graphite at temperatures above 100 K

**Authors:** S. Huberman[1]†, R. A. Duncan[2]†, K. Chen[1], B. Song[1], V. Chiloyan[1], Z. Ding[1], A. A. Maznev[2], G. Chen[1]*, K. A. Nelson[2]*.

**Affiliations:**

[1] Department of Mechanical Engineering, Massachusetts Institute of Technology, Cambridge, Massachusetts 02139, USA.

[2] Department of Chemistry, Massachusetts Institute of Technology, Cambridge, Massachusetts 02139, USA.

*Correspondence to: gchen2@mit.edu, kanelson@mit.edu

†These authors contributed equally to this work.

**Abstract:**

Wavelike thermal transport in solids, referred to as second sound, has until now been an exotic phenomenon limited to a handful of materials at low temperatures. This has restricted interest in its occurrence and in its potential applications. Through time-resolved optical measurements of thermal transport on 5-20 μm length scales in graphite, we have made direct observations of second sound at temperatures above 100 K. The results are in qualitative agreement with *ab initio* calculations that predict wavelike phonon hydrodynamics on ~ 1-μm length scale up to almost room temperature. The results suggest an important role of second sound in microscale transient heat transport in two-dimensional and layered materials in a wide temperature range.

**One Sentence Summary:**

Wavelike thermal transport is observed at above 100 K and predicted at even higher temperatures, suggesting prospects for unique microscale cooling kinetics in two-dimensional and layered materials.

**Main Text:**

In nonmetallic solids, heat is carried by lattice vibrations, or phonons. In nearly perfect crystals at temperatures below ~10 K, phonons may propagate over macroscopic distances without scattering, yielding ballistic heat transport. At room temperature, the mean free paths of heat-carrying phonons are short due to high rates of phonon-phonon scattering, and heat spreads macroscopically by diffusion. In recent years, the transition between ballistic and diffusive transport kinetics (transport distance $D$ and time $t$ related by $D \propto t$ or $D \propto \sqrt{t}$ respectively) on micro- and nano-scales has attracted increased interest [1,2], leading to observations of non-diffusive heat transport at room temperature in silicon and other materials [3,4,5].

Second sound, or the wavelike transport of heat in solids [6] has long been considered an exotic phenomenon occurring at very low temperatures in but a few materials in a special regime between ballistic transport and diffusion. In this regime, referred to as phonon hydrodynamics [7,8], "normal" phonon-phonon scattering processes that conserve the net reduced phonon momentum are much more frequent than "umklapp" processes capable of reversing the reduced momentum direction. Normal scattering alone cannot dissipate a heat flux and return the lattice to thermal equilibrium; instead, the phonon population relaxes to a "displaced" Bose-Einstein distribution characterized by a nonzero drift velocity [9], akin to a flow of molecules in a gas. This enables temperature waves (i.e., phonon density waves) propagating at a velocity below the speed of sound, which have been termed second sound in analogy with temperature waves in superfluid He [10].

Second sound in solids has been observed in heat pulse experiments in $^3$He in a temperature range of 0.42 to 0.58 K [11], in Bi between 1.2 and 4 K [12], and in NaF between 11 and 14.5 K [13,14]. The occurrence of second sound in $SrTiO_3$ at 30-40 K has been reported [15,16]; however, other authors argued that the interpretation of the low-frequency doublet seen in Brillouin spectra in terms of second sound is not supported by the observations [17,18]. More recently, it has been suggested that the temperature window for phonon hydrodynamics is much wider in graphene [9] and other 2D materials [19] due to strong normal scattering involving the flexural mode. A similar prediction has been made for graphite, which exhibits a phonon dispersion resembling that of graphene due to the weakness of its interlayer van der Waals interactions [20].

In this paper, we report that second sound is indeed observed in graphite at temperatures over 100 K. We employ impulsive optical excitation to set up a spatially sinusoidal temperature profile in the sample and observe dynamics unambiguously showing wave-like behavior. In contrast to the previous observations in macroscopic samples at very low temperatures, second sound in graphite is only observed on the microscale (with a characteristic propagation length on the order of 10 μm) and is found to propagate faster than the slow transverse acoustic waves (but slower than the fast transverse mode). The observations are generally consistent with *ab initio* numerical simulations.

Second sound is expected to occur on the time scale that falls in between the characteristic normal and Umklapp relaxation times—i.e., $\tau_N < t < \tau_U$. Theoretical predictions for graphene [9] indicated that experiments should be done on the nanosecond scale, making it difficult to use conventional thermal sensors to probe the transport. Time-resolved laser-based measurements, on the other hand, are well suited for measuring thermal responses on this timescale *(1,3,4,5)*. We used the transient thermal grating (TTG) technique (Fig. 1A), in which two short (60 ps) laser pulses are crossed at the surface of the sample, yielding a spatially sinusoidal heat source whose period *L* is set by the optical interference pattern. This heat source sets up a "thermal grating", i.e., a spatially sinusoidal temperature field along the surface, $\Delta T(t,z)\cos(qx)$, where $q=2\pi/L$ is the TG wavevector; this thermal grating subsequently decays via thermal transport. Thermal expansion produces an associated sinusoidal surface displacement modulation or "ripple" pattern $u(t)\cos(qx)$ that acts as a transient diffraction grating for probe laser light. The decay of the thermal grating due to thermal transport is thus monitored through the time-dependent diffraction of a continuous-wave probe laser beam. The diffracted beam is superposed with a reference beam derived from the same laser source for optical heterodyne detection *(21)*. The signal is proportional to the amplitude of the spatially sinusoidal surface displacement modulation $u(t)$ *(22)*.

Our sample was highly oriented pyrolytic graphite of the highest quality commercially available grade. The sample was polycrystalline, with grain size ~10 μm *(22)*, but the *c* axes of all crystal grains were perpendicular to the sample surface. The absorption of 515 nm excitation light produced the initial thermal grating within an optical skin depth of ~30 nm. Thermal transport occurs both along the thermal grating (in-plane) and in the direction normal to the surface (cross-plane). However, the cross-plane thermal conductivity of graphite is ~300 times lower than the in-plane conductivity, hence the cross-plane thermal diffusion depth remains much smaller than the TG period until the TG is washed out by in-plane transport *(22)*. In this case, the normal surface displacement is proportional to the integral of the temperature rise over the cross-plane coordinate *z*. Consequently, the cross-plane transport does not affect the surface displacement, and the measured signal is only sensitive to one-dimensional (1D) in-plane transport in the grating direction *x*. According to the heat diffusion equation, in the 1D case a TTG decays exponentially with a time constant $\tau = L^2/4\pi^2\alpha$, where *L* is the grating period and $\alpha$ is the thermal diffusivity *(23)*. Indeed, at room temperature (300 K) we observe exponential decay of the TTG as shown in Fig. 1B. At *L* = 37.5 μm, we obtain a thermal diffusivity of 11 cm²/s, which agrees with the literature *(24)*. However, as the TTG period is reduced, the quantity $L^2/4\pi^2\tau$, i.e., the apparent thermal diffusivity, does not remain equal to the constant value of $\alpha$ as predicted by the heat diffusion equation. As shown in Fig. 1C, at small TTG periods the decay becomes slower than the diffusion model predicts, indicating nondiffusive behavior similar to that observed previously in Si and other materials *(3,4,5)*. What we are seeing is the onset of the size effect occurring when the mean free path of heat-carrying phonons becomes comparable to the heat transport distance *(25)*. Still, the decay remains exponential over the whole range of TTG periods used, i.e. 3.7 – 37.5 μm.

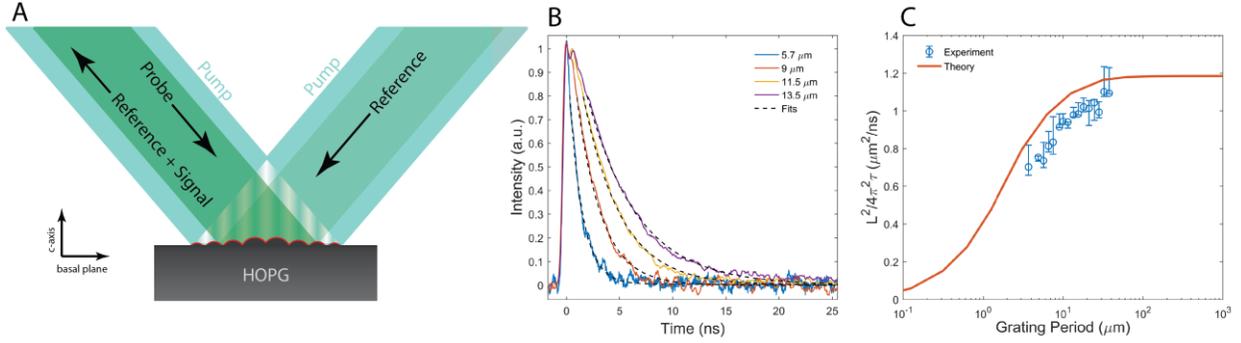

**Fig. 1.** (A) Schematic illustration of the experiment. A thermal grating is produced by two crossed pumped pulses; thermal expansion gives rise to a surface modulation that is detected via diffraction of the probe beam; the reference beam is used for optical heterodyne detection. (B) Signal waveforms at 300 K for a range of TTG periods. The dashed lines correspond to the best fits to exponential decays. (C) The measured normalized TTG decay rate $L^2/4\pi^2\tau$ as a function of grating period $L$ at 300 K for grating periods of 3.7-37.5 μm, plotted alongside theoretical results from *ab initio* BTE calculations. In the diffusive regime this quantity would be a constant, equal to the thermal diffusivity $\alpha$. The reduction of the apparent diffusivity as $L$ is reduced occurs as the mean free paths of increasing fractions of heat carrying phonons become comparable to the grating period.

When the temperature is lowered to 85 K, strikingly different behavior is observed, as shown in Fig. 2A. Unlike the signals at 300 K which decay monotonically, signal waveforms at 85 K yield damped oscillations, with the signal falling below zero *(26)*. In the heterodyne detection scheme, this sign flip of the TTG signal means that the spatial phase of the grating has shifted by $\pi$ *(21)*, i.e. the local maxima of the surface displacements (and hence of the temperature) are located where the local minima used to be, and vice versa; in other words, the TTG behaves as a thermal standing wave. The sign flip is a hallmark of the wave-like propagation of heat: in the diffusive transport regime, TTG maxima and minima cannot switch places because heat can only move from hotter to colder regions. With increasing TG period, the negative dip in the response becomes shallower and eventually disappears. The position of the dip shifts to longer times as the period increases, indicating that the frequency of the wave-like dynamics decreases. The inset in Fig. 2A shows this frequency, determined from the position of the first minimum of the response (corresponding to ½ of the oscillation period) as a function of the wavevector $q$. The nearly linear dependence indicates a velocity of about 3200 m/s (determined from the slope of the linear fit multiplied by $2\pi$). It should be noted that TTG signals often contain oscillations due to surface acoustic waves (SAWs), albeit with much lower damping rates *(27)*. However, the observed oscillation frequency does not match the frequency of SAWs or any other acoustic waves that may propagate in the basal plane of graphite: the SAW velocity is 1480 m/s *(28)*, which is very close to the slow transverse velocity, whereas the fast transverse velocity is 14700 m/s and the longitudinal velocity is even higher *(29)*. Besides, there would be no reason for acoustic waves to disappear as the background temperature or the grating period are increased.

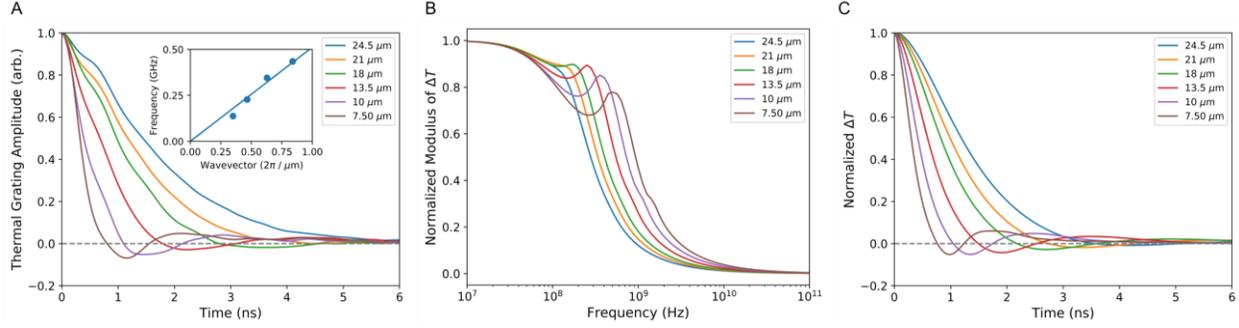

**Fig. 2.** (A) TTG signal at 85 K for a range of grating periods. In the inset, circles represent the measured second sound frequency as a function of the wavevector and the solid line is a linear fit corresponding to a phase velocity of 3200 m/s. (B) Absolute value of complex frequency-domain Green's functions vs frequency at 80 K for a range of TTG periods. (C) Simulated thermal grating amplitude vs time at 80 K.

To simulate the observed dynamics, we solved the linearized Boltzmann transport equation (BTE) with the full three-phonon scattering matrix in the one-dimensional TTG geometry, i.e. for the initial temperature profile $\Delta T_0 \cos(qx)$ *(22)*. Rigorous solutions of the BTE for graphene and graphite were previously used to calculate the thermal conductivities of these materials *(30,31)* and to explore the conditions for the phonon hydrodynamics regime *(9,19,20)*. However, previous studies dealt with the stationary BTE which is unable to capture transient phenomena such as second sound. Recently, Chiloyan et al. *(32)* described a technique for calculating frequency-domain Green's functions for the non-stationary and spatially non-uniform BTE. Such Green's functions, illustrated in Fig. 2B describe a response of the phonon population to a heat source having the form of a harmonic plane wave, $\exp(i\omega t - \mathbf{qr})$. In our case, the heat source is sinusoidal in space and can be modeled as impulsive in time since the laser pulse duration is short compared to the observed dynamics. Consequently, temporal dynamics of the TTG can be obtained by performing a Fourier transform of the frequency-domain Green's function at wavevector *q* corresponding to the TTG period. We follow the approach developed by Chiloyan et al. to solve the BTE with a collision integral constructed using inputs from density functional theory calculations, accounting for three-phonon scattering as well as scattering by mass disorder due to the natural isotope content (1.1% of $^{13}C$). Our calculation, performed entirely from first principles with no fitting parameters, produces the time dependence of the thermal grating amplitude which can be directly compared to the experimentally measured signal.

Calculations performed at 300 K over a range of grating periods yielded exponential TTG decays in agreement with the experiment. The decay times of the simulated waveforms yielded the dependence of the apparent diffusivity on the TTG period reproducing the trend seen in the experiment, as shown in Fig. 1C. In contrast, at 85 K the frequency-domain Green's functions yield a resonant peak as seen in Fig. 2B. This peak is a hallmark of second sound and gives rise to damped oscillations in the simulated time-domain waveforms shown in Fig. 2C. The simulated waveforms agree qualitatively with the experimental data (see Supplementary material for a detailed comparison of measured and simulated waveforms) and yield trends with respect to the TTG period that are consistent with the observations, including the disappearance of the second sound signature at large TTG periods. The calculated second sound velocity determined from the peak position of the frequency resonance in Fig. 2C is 3650 m/s, which is somewhat higher than the measured velocity. The theory correctly predicts that the second sound velocity falls in between the slow and fast transverse acoustic velocities. In contrast to graphite, in other materials second sound was found to be slower than the slowest phonon velocity. The peculiarity of graphite is the exceptionally low velocity of the slow transverse acoustic mode, which is analogous to the flexural ZA mode in 2D materials such as graphene. It is the extremely large anharmonicity and density-of-states of this mode that lead to intense normal scattering and create conditions for hydrodynamic phonon transport *(9)*.

Figure 3 shows TTG data with a constant grating period of 10 μm at different temperatures. The oscillatory behavior is still apparent at 104 K and even at 125 K, and eventually disappears at 150 K. This trend is well reproduced by the simulations. The oscillatory behavior also disappears when the temperature is lowered to 50 K. At temperatures below 80 K we see increasing discrepancy between experiment and simulations: at 50 K the simulated response still contains a dip at ~1.5 ns, even though it no longer goes negative. One possible origin of the quantitative discrepancy is the assumption of the initial thermal distribution used in our simulations *(22)*. This assumption becomes increasingly inaccurate at low temperatures when transport transitions to the ballistic regime. One would need to consider the details of the electronic excitation by the pump laser pulse and the subsequent electron-phonon interaction to determine the initial phonon distribution in the laser-induced thermal grating. Another pertinent consideration is that the experimental observable is thermal expansion whereas the simulated observable is the temperature (i.e., thermal energy per unit volume). While in equilibrium thermal expansion should precisely track the temperature, in a non-equilibrium situation this does not have to be the case since the Grüneisen parameter is different for different phonon modes. Figure 3 also shows a simulated response at 50 K obtained in the ballistic limit, i.e., with phonon scattering rates set to zero. One can see that the dip in the response is expected to disappear in the ballistic regime. Thus the observed disappearance of the second sound signature at 50 K is generally consistent with what is expected to happen in the transition to the ballistic regime. Interestingly, eliminating scattering altogether makes the TTG decay slower: normal phonon-phonon scattering processes facilitate heat transport by redistributing energy to modes with higher group velocity.

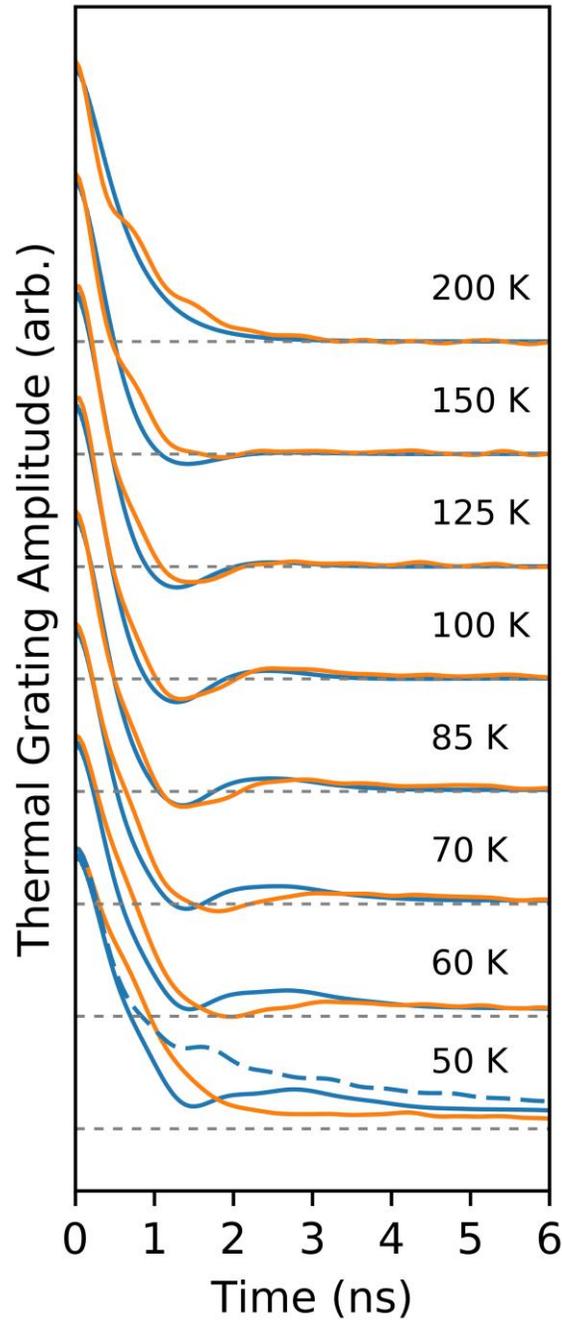

**Fig. 3.** (A) Measured TTG signals (orange curves) and simulated responses (blue solid curves) at different temperatures. (The label 85K corresponds to the experimental curve, while the simulated curve was obtained for 80 K). Horizontal dashed lines indicate zero for each pair of curves. The dashed curve for 50 K shows the calculated response in the ballistic limit, i.e., with the scattering matrix set to zero.

It is instructive to determine the second sound domain in the temperature–TTG period parameter space. We use the ratio of the maximum at the peak of the magnitude of frequency-domain Green's function (Fig. 2B) and the minimum between the peak and zero frequency as a metric for the strength of the second sound effect. As one can see in Fig. 4, second sound is predicted to occur between 50 and 250 K, with higher temperatures corresponding to shorter thermal transport length scales: while for $L$=10 μm the temperature window closes at ~150 K in agreement with our observations, at $L$=1.5 μm a second sound signature is expected to be observable up to 250 K. (Our experimental setup lacked the temporal resolution needed to probe responses at TTG periods smaller than ~5 μm.) At low temperatures and small grating periods, phonon scattering of any kind disappears and transport becomes ballistic. At high temperatures and large TTG periods, transport slowly transitions to the "quasi-diffusive" regime illustrated in Fig. 1B,C (TTG decay is exponential, but the decay time $\tau$ does not scale as $L^2$), and finally the diffusive limit ($\tau \propto L^2$) would be reached.

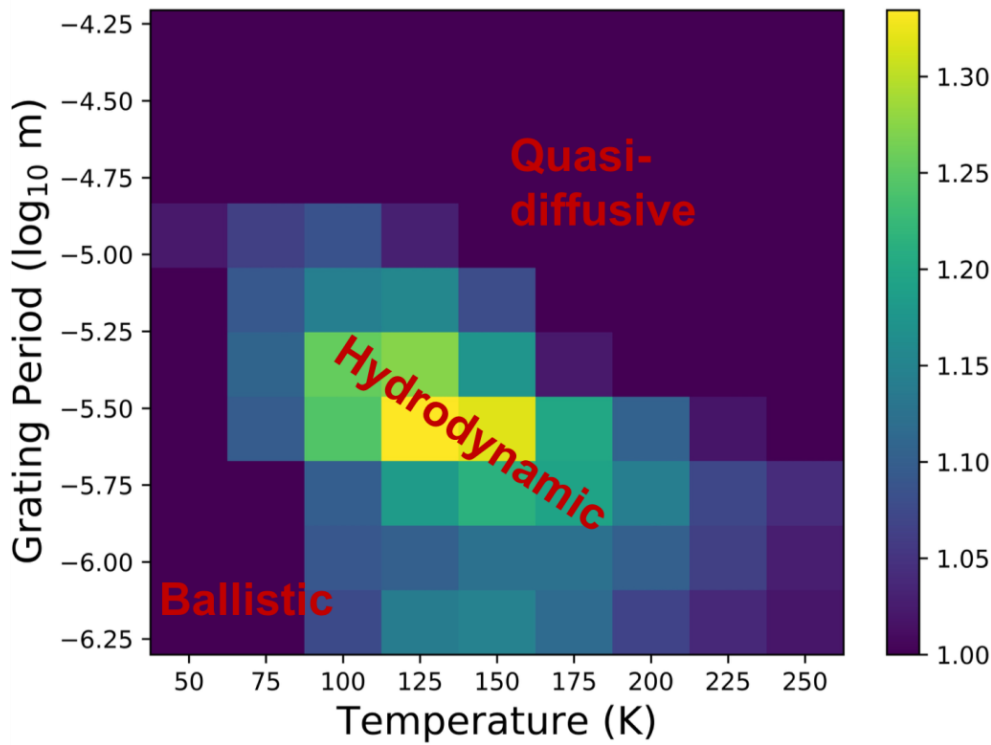

**Fig. 4.** Second sound window for graphite with the natural isotope content. The color scale corresponds the ratio of the maximum at the peak and the minimum between the peak and zero frequency in the frequency domain responses such as those shown in Fig. 2B.

The size of the second sound window depends on the concentration of defects, such as isotopes, which contribute to umklapp scattering *(6,9,20)*. While the second sound window in Fig. 4 is shown for the natural carbon isotopes content, it is expected to be wider for isotopically pure material *(9)*. Previously, second sound was only observed in isotopically pure solids (with the exception of the controversial $SrTiO_3$ reports). The fact that second sound can be seen in graphite with the natural isotope content indicates the unique nature of phonon hydrodynamics in this type of material.

In summary, we have experimentally observed second sound in graphite at temperatures 85-125 K using the TTG technique, in agreement with *ab-initio* predictions based on solving the full three-phonon scattering matrix BTE. The significance of this observation is not only in that second sound is observed at unusually high temperatures. While previous observations of this phenomenon have been conducted on the macroscale with extremely pure single crystal samples, we observed second sound on the microscale in a polycrystalline sample with isotopic disorder. Consequently, second sound and other phonon hydrodynamic effects in graphite, graphene, and, possibly, other layered and 2D materials can no longer be considered exotic phenomena with little practical consequence. We believe that phonon hydrodynamics may very well have an impact, for example, on the performance of graphite and graphene as heat spreading materials in microelectronics. Furthermore, our work complements recent observations of electronic hydrodynamic transport in graphene and other 2D materials *(33,34,35)*. We expect these developments to open up exciting potential for understanding and manipulating transport phenomena on micro- and nano-scales.

**Acknowledgments:** We are grateful to Lorenzo Paulatto for his help with the construction of the scattering matrix.

**Funding:** This work is supported in part by the office of Naval Research under MURI grant N00014-16-1-2436 (G.C. for high thermal conductivity materials including phonon hydrodynamics) in part by the Solid State Solar-Thermal Energy Conversion Center (S$^3$TEC), an Energy Frontier Research Center funded by the U.S. Department of Energy, Office of Science, Office of Basic Energy Sciences under Award DE-SC0001299 (G.C. for thermoelectric materials), and in part by the National Science Foundation EFRI 2-DARE Grant EFMA-1542864 (K.A.N. for thermal transport in 2D materials).

**Author Contributions**: The project was proposed by S.H. and R.A.D. who led the theoretical and experimental work respectively, with a significant experimental contribution by K. C. and B. S. and theoretical contributions by V.C., and Z.D, and with guidance from A.A.M., G.C., and K.A.N. All the participants contributed to the writing of the paper.

# Supplementary Materials for

## Observation of second sound in graphite at temperatures above 100 K

**Authors:** S. Huberman[1]†, R. A. Duncan[2]†, K. Chen[1], B. Song[1], V. Chiloyan[1], Z. Ding[1], A. A. Maznev[2], G. Chen[1]*, K. A. Nelson[2]*.

Correspondence to: gchen2@mit.edu, kanelson@mit.edu

**Materials and Methods**

Experimental Methods

The experimental design for the TTG measurements reported in this work has been described in detail elsewhere *(36)*. In short, the ±1 orders of a 515 nm pulsed pump laser source diffracted from a binary transmission grating "phase mask" are imaged onto the sample through a two-lens telescope system, where absorption of the resulting sinusoidal intensity profile establishes the transient thermal grating. The ±1 orders of a 532 nm continuous probe beam diffracted from the same spot on the phase mask are similarly collected and sent through the same two-lens telescope system, one order becoming the probe beam that diffracts from the time-evolving transient grating and the other order becoming the reference beam which is superposed with the diffracted signal to achieve heterodyne detection. The reference beam is attenuated with a neutral density filter, and the relative phase between the probe and reference beams is controlled by tilting a highly-parallel optical flat through which the probe beam passes prior to impinging on the sample. The heterodyne phase *(21,36)* was calibrated using a reference sample (tungsten) yielding a SAW signal corresponding to a pure phase grating. Using this calibration, we established that the amplitude grating (thermoreflectance) thermal signal from graphite is absent and we only observe a phase grating response, consistent with the assumption that the signal comes from surface displacement. The dependence of the acquired waveform on the heterodyne phase $\theta$ is shown in Fig. S1, where the thermal decay is seen to be entirely in the phase grating. All time-domain data reported in this paper are obtained by subtracting the traces acquired at $\theta = -\pi/2$ by traces acquired at $\theta = \pi/2$ in order to eliminate any background signal that does not correspond to a material excitation at the TTG wavevector. All measurements were performed in a cryostat under vacuum (with pressure not exceeding 1 Torr).

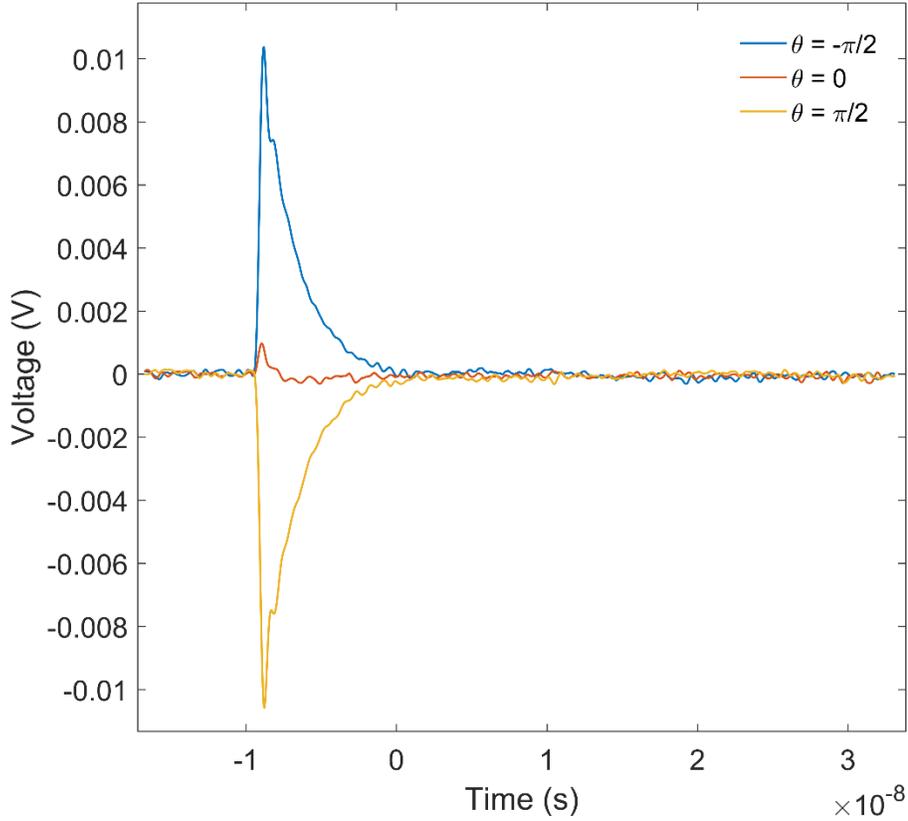

**Fig. S1.** Heterodyne phase ($\theta$) dependence of the acquired waveform at a grating period of 10 μm and a temperature of 300 K. $\theta = \pm\pi/2$ is taken to correspond to pure phase grating signal. At $\theta = 0$ the thermal decay entirely disappears. This indicates that the material response to the TTG acts as a transient phase grating, consistent with the observed decay arising solely from thermal expansion-driven surface displacement.

The room-temperature data shown in Figs. 1B and 1C were obtained with pump, probe, and reference spot sizes of 200 μm in diameter ($1/e^2$ intensity). The repetition rate of the pump was 1 kHz, and the acquisition rate of the oscilloscope was 300 Hz. The pump pulse duration was 60 ps, and the probe beam was modulated by an electrooptical modulator with a 5% duty cycle. The data corresponding to grating periods of 4.9 μm and larger were acquired using a Hamamatsu photodetector (C5658) with a 1 GHz bandwidth, whereas the traces taken at grating periods smaller than 4.9 μm were acquired using a New Focus photodetector (1591NF) rated with a 3.5 GHz bandwidth. The pump pulse energies for all traces at room temperature were within a range of 85-170 nJ, and the average probe powers at the sample were within a range of 2-8 mW (i.e., peak probe powers between 40 – 160 mW).

The low-temperature measurements shown in Figs. 2A and 3 were performed with a pump spot size of 168 μm in diameter and a pump pulse duration of 15 ps. The probe and reference beams both had spot sizes of 135 μm in diameter, and unlike the room-temperature measurements the probe was not electro-optically modulated. The repetition rate of the pump and the mean acquisition rate of the oscilloscope were 2 kHz and 1.8 kHz, respectively. Temperature was controlled by cryogenic cooling and regulated heating with a resistive heater. Liquid $N_2$ (He) was used as the cryogen for all measurements performed at temperatures higher (lower) than 80 K. The pump pulse energy was 8 -10 nJ and the probe power was 60 mW.

Since the heating of the sample by the pump and/or probe laser light may increase the background temperature and thus affect thermal transport measurements, special care was taken to quantify the effect of the laser powers on the measurements. For the room temperature measurements shown in Fig. 1B, three traces were taken for each grating period: one at a baseline set of pump and probe powers, one at a doubled pump power and baseline probe power, and one at a baseline pump power and doubled probe power. From these triplets the systematic error associated with power effects was calculated as the difference between the exponential decay times obtained at baseline and high-power traces. The points in Fig. 1C were calculated from the decay times measured at the baseline laser powers, and the error bars show the uncertainties associated with laser power effects and statistical uncertainty. The effective thermal diffusivity values shown in Fig. 1C were obtained by fitting the raw TTG data over a domain beginning at the time at which the signal had decayed to 80% maximum amplitude. The amplitude of the thermal grating at the beginning of the fitted time window and the average temperature increase due to the pump for the room-temperature measurements were estimated to be at most 77 K and 48 K, respectively, and the heating due to the probe beam was estimated to be at most 4 K.

At an ambient temperature of ~85 K the temperature grating amplitude immediately upon pump incidence was estimated to be 57 K and the average temperature rise due to pump incidence was estimated to be 28 K. However, by the time the second sound signature is observed substantial thermal transport in both cross- and in-plane directions has occurred, significantly reducing these values. Assuming that the heat equation is still valid for cross-plane transport, the cross-plane thermal diffusion length at 85 K corresponding to the time at which the second sound dip reaches its minimum (i.e., ~1.2 ns for the 7.5 µm trace) is about 560 nm. Given that the optical penetration depth of graphite is ~30 nm for 515 nm light, by this time cross-plane transport has reduced the surface temperature by a factor of ~20 relative to the temperature immediately after pump incidence. Thus, by the time the second sound signature is observed the average temperature is estimated to be ~1.4 K higher than the ambient 85 K. Furthermore, in-plane transport further reduces the temperature grating amplitude by a factor of the normalized value of the TTG signal. The maximum magnitude of the dip below the baseline is ~7% the maximum amplitude of the TTG signal, yielding a maximum grating amplitude of ~0.2 K during the second sound oscillation. The heating from the probe beam was estimated to be 0.9 K.

We verified that changing either the excitation energy or probe power by a factor of two has no effect on the shape of the measured signal waveforms at 85 K. We had to increase the excitation energy by as much as a factor of 10 to see a noticeable change in the normalized signal, see Fig. S2.

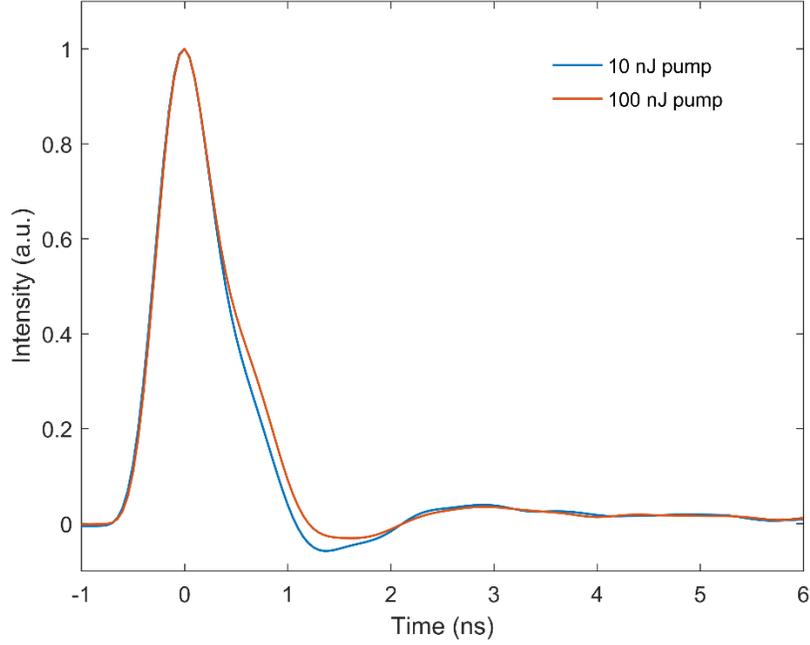

**Fig. S2.** Measured TTG signals at 85 K and a period *L*=10 µm for pump pulse energies 10 and 100 nJ normalized to unity at the maximum. We had to increase the pump energy by a factor of 10 to see an appreciable change in the normalized waveform.

Numerical Methods

The TTG signal was numerically simulated using the Green's function technique for the BTE with the full scattering matrix developed by Chiloyan *et al.* *(32)*. We begin with the linearized BTE in the deviational form:

$$\frac{\partial g_i}{\partial t} + \vec{v}_i \cdot \vec{\nabla} g_i = Q p_i + \sum_j W_{ij}(g_j^0 - g_j) \tag{S1}$$

Here $g_i$ are the mode-specific deviational energy densities and $g_i^0$ are the equilibrium energy densities at the local temperature per unit volume of the Brillouin zone, $\vec{v}_i$ are the group velocities, $Q$ is the heat density generation rate per unit volume, $p_i \equiv c_i/C$ for mode specific heat $c_i$ and total specific heat $C \equiv \sum_i c_i$, and $W_{ij}$ is the full scattering matrix that accounts for scattering events due to anharmonic couplings between phonon populations and scattering due to isotopic disorder:

$$W_{ij} = W_{ij}^{3-phonon} + W_{ij}^{isotope} \tag{S2}$$

The exact expressions for the matrix elements of $W$ are found in Ref. 37 (see Eqs. 5, 6, 9 and 13). Briefly, the 3-phonon scattering amplitudes are related to the third-order force constants of the interatomic potential and isotope disorder is modeled as a mass-difference perturbation.

We consider a one-dimensional TTG geometry with a heat source

$$Q = \cos(2\pi x/L)\delta(t) \tag{S3}$$

creating an initial sinusoidal temperature distribution. Even though in the experiment the thermal grating is not one-dimensional, we believe that this is a reasonable approximation, since transport in the z direction does not affect the experimental observable (i.e. the surface displacement amplitude). The choice of factor $p_i$ in Eq. (S1) is based on the assumption that the source generates a thermalized phonon population.

Linearizing Eq. (S1) in temperature deviation $\Delta T$, taking the spatial and temporal Fourier transforms, and inverting the resulting algebraic equation yields an explicit equation for the deviational energy densities:

$$\tilde{g}_i = \tilde{Q} A_{ij}^{-1} p_j + \Delta\tilde{T}\big(\delta_{ik} - i[A^{-1}]_{ij} D_{jk}\big) c_k \tag{S4}$$

where tildes denote Fourier transformed variables over both time and space, $A_{ij} \equiv W_{ij}\frac{\omega_i}{\omega_j} + i\delta_{ij}(\omega + \vec{q}\cdot\vec{v}_i)$, $D_{ij} \equiv \delta_{ij}(\omega + \vec{q}\cdot\vec{v}_i)$, $\omega_i$ are the phonon frequencies, and $\vec{q}$ and $\omega$ are the Fourier domain variables from the Fourier transforms over space and time respectively. Expressing the temperature deviation as

$$\Delta\tilde{T} = \frac{1}{C}\sum_i \tilde{g}_i \tag{S5}$$

and using Eq. (S3) and Eq. (S4) one obtains

$$\Delta\tilde{T}(q,\omega) = \tilde{Q}\frac{\sum_i [A^{-1}\vec{p}]_i}{\sum_i [iA^{-1}D\vec{c}]_i}. \tag{S6}$$

For the one-dimensional transient grating geometry imposed by Eq. (S3) the solution only needs to be obtained for $q = 2\pi/L$. To obtain the simulation of the time-domain transient, we perform an inverse Fourier transform of the solution given by Eq. (S6) over $\omega$.

The effective thermal conductivity of the steady state 1D thermal grating is given by

$$k_q = \frac{1}{q^2}\frac{\sum_i [iA^{-1}D\vec{c}]_i}{\sum_i [A^{-1}\vec{p}]_i} \tag{S7}$$

where the matrix A is constructed with $\omega = 0$ *(31)*. This calculation is valid in the quasi-diffusive regime where the decay remains exponential and was used, upon dividing by the heat capacity, to obtain the theoretical apparent thermal diffusivities of Fig. 1C.

The second- and third-order force constants for graphite calculated by Ding *et al. (20)* were used as inputs to construct $W_{ij}$ on a 16 x 16 x 8 wavevector mesh with a Gaussian smearing parameter of 20 cm$^{-1}$ as an approximation for the Dirac delta function. The $W_{ij}$ matrix was constructed using the D3Q module of Quantum-Espresso *(37)*. The matrices involved in this computation were on the order of 25,000 x 25,000 values in size, taking approximately 30 minutes to directly invert on a 2.8 GHz CPU core. Matrix inversion was performed using the Numpy package and a trivial parallelization scheme where Eq. (S5) for each tuple of ($\omega, \vec{q}$) was calculated independently.

**Supplementary Text**

Characterization of the sample

The sample investigated in this study was highly-oriented pyrolytic graphite purchased from SPI supplies (SPI HOPG Grade 1). According to the manufacturer, the mosaic spread of the sample is $0.4° \pm 0.1°$. Electron back-scatter diffraction (EBSD) was performed to analyze the in-plane grain structure. The results of this analysis are shown in Fig. S3, where typical sizes of grains are found to be ~10 μm.

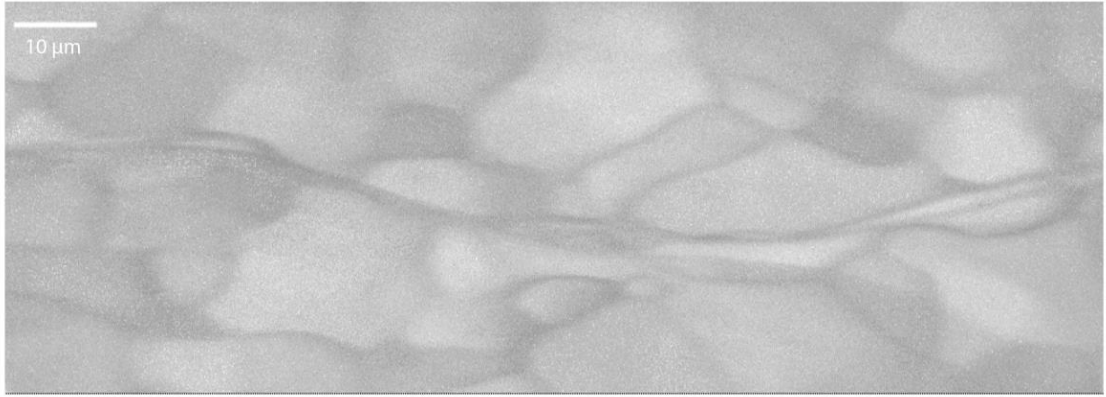

**Fig. S3.** EBSD image quality (IQ) map of the sample surface showing the sample's grain structure in the basal plane.

Analysis of the detector response

To determine the detector response, we measured the response to a pump pulse 60 ps in duration, which approximates a delta function relative to the time scale of the measurement. The detector response and the experimental TTG results at 300 K and a grating period of 11.5 μm are shown in Fig. S4A. Also plotted is an exponential function with a step-like rise at $t = 0$ and the convolution of this exponential with the detector response. We see that the convolved waveform has a rise that agrees well with the experiment, and features small-amplitude oscillations that correspond well with the oscillations observed (the experimental trace does not match the convolved exponential when normalized by maximum amplitude due to early-time nonequilibrium dynamics, but the observed oscillations are nevertheless reproduced). Thus, these small oscillations in the room-temperature data are detection artifacts that arise from the fast rise time of the TTG signal, and do not correspond to a material response.

The second sound signature in the low-temperature measurements are also oscillations on a nanosecond timescale, and one might worry that these too are detection artifacts rather than a material response. Figure S4B shows the TTG signal measured at a temperature of 85 K and a grating period of 10 µm and a (steplike) exponential waveform with a similar decay time as well as their convolutions with the measured detector response. The convolved exponential waveform exhibits small ripples with the same periodicity as the oscillations in the detector response, but it does not recreate the observed dip below the baseline. If the measured signal is itself convolved with the detector response the resulting waveform remains virtually unchanged. These observations suggest (1) that a convolution with the detector response does not mimic the observed second sound signature, and (2) that the effect of the detector response on the measured TTG signal is minimal. Furthermore, in Fig. S4C, the measured trace at 85 K and 10 µm TTG period is plotted alongside a trace at 300 K and 3.7 µm period measured with the same Hamamatsu C5658 photodetector. We see that the two traces are of a similar timescale, and indeed a weak effect of the detector response is evident in the 300 K trace in the form of a weak half-cycle modulation with the periodicity of the oscillations seen in the detector response (similar to those seen in the convolved exponential waveform in Fig. S4B; it is also very slightly evident in the 85 K trace in Fig. S4C as well). However, the measurement at 300 K does not exhibit the sub-baseline dip observed in the 85 K measurement, indicating that this feature in the 85 K trace cannot be the result of the detector response.

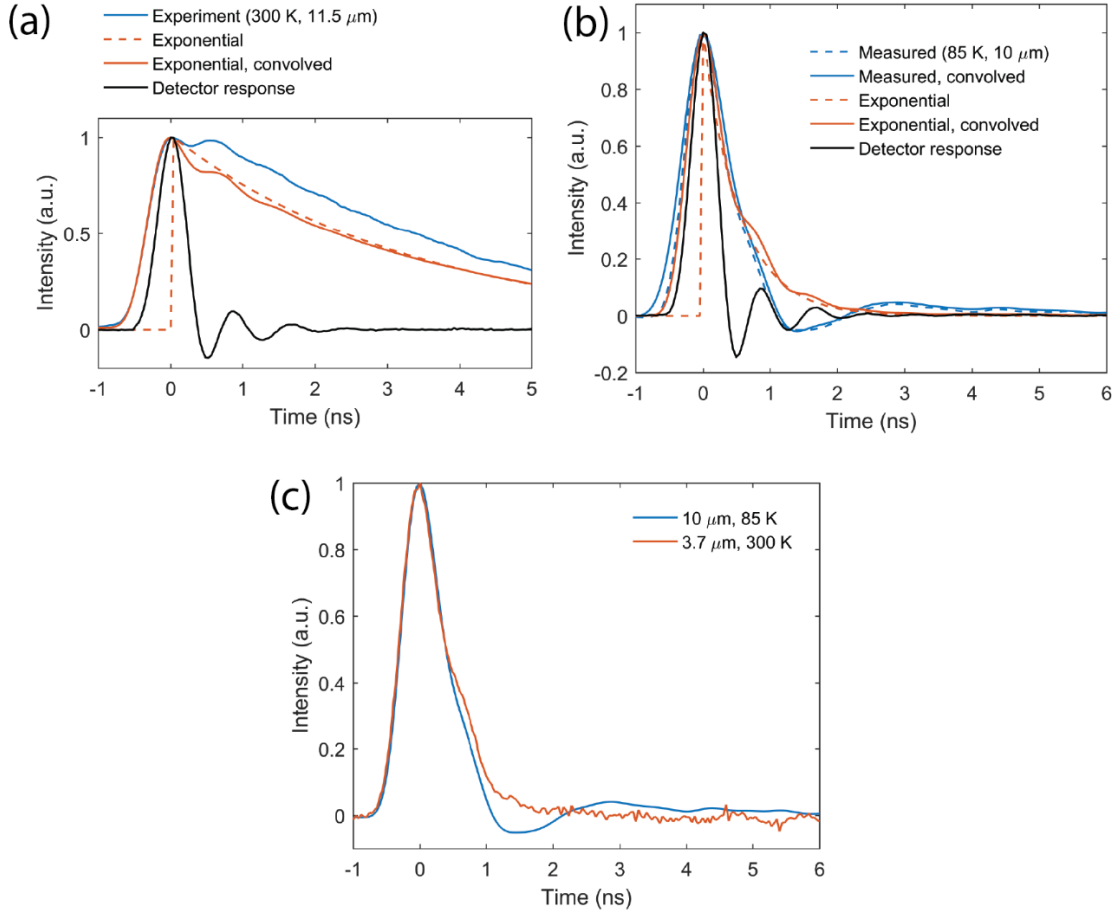

**Fig. S4.** (a) Measured detector response alongside room-temperature data with a grating period of **11.5 µm**. An exponential with a step-like rise and its convolution with the detector response are also plotted, yielding small oscillations that are consistent with those observed in the room-temperature data. (b) Measured TTG signal at 85 K

and a period of 10 µm and an exponential waveform with a similar timescale, both plotted alongside their convolutions with the measured detector response. The convolved exponential waveform does not yield a sub-baseline dip, and the effect of convolution on the measured waveform is minimal. (c) TTG measurements at 85 K with a 10 **µm** period and at 300 K with a 3.7 **µm** period. These two traces are very close in timescale, but only in the low-temperature trace is the oscillation observed. This indicates that the oscillation is not a detector response artifact.

Comparison of individual experimental waveforms with simulations

Figs. S5 is a reproduction of the data plotted in Figs. 2A,C of the main text, but now each measured waveform is plotted separately alongside the corresponding simulated waveform.

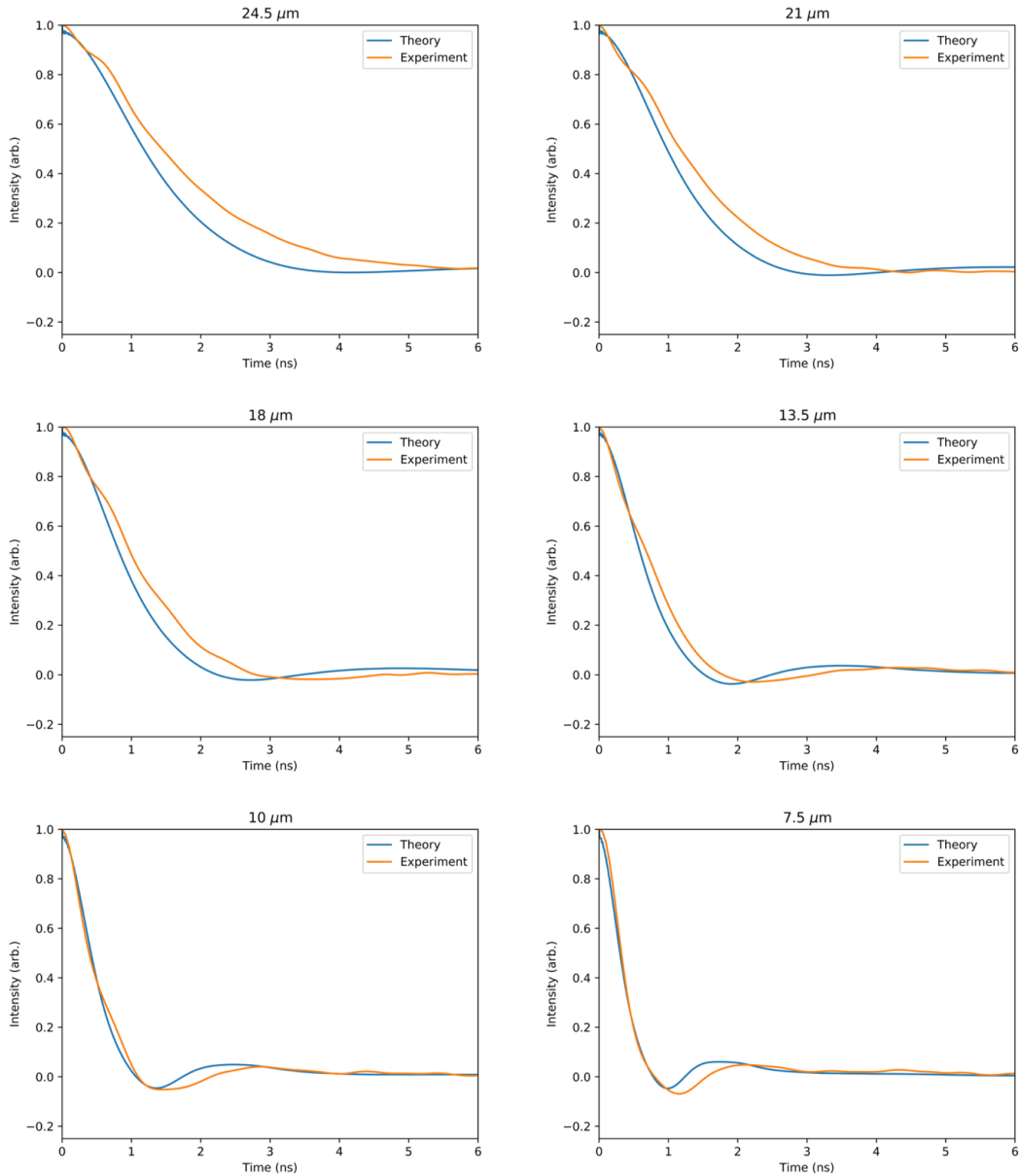

**Fig. S5.** Experimental and numerical results at 85 K over a range of grating periods. Reproduction of traces shown in Figs. 2A,C.

Isotope effect on second sound

To investigate the role of isotopes, we performed additional BTE calculations for isotopically-pure graphite (where $W_{ij}^{isotope} = 0$) at 100 K for a 10 µm period grating, which are reported in Fig. S6. One can see that the second sound damping is significantly reduced in the absence of isotope mass-disorder scattering.

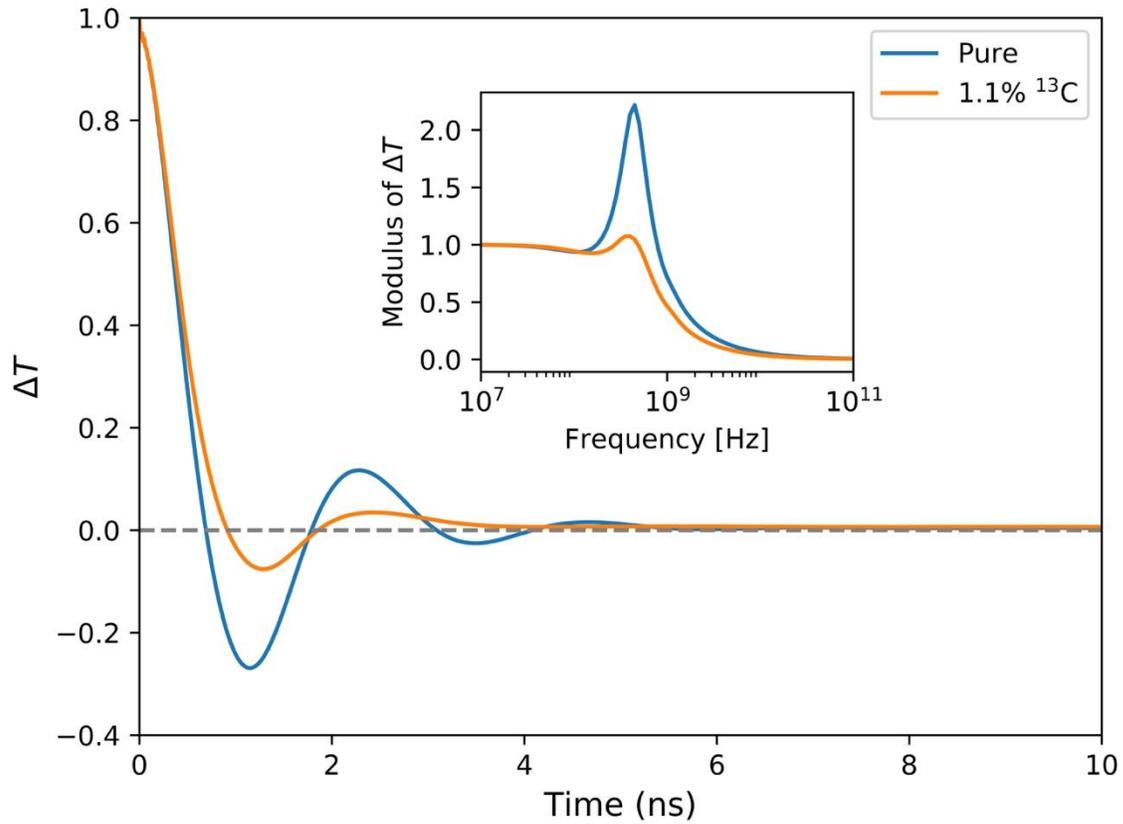

**Fig. S6.** Time-domain calculated temperature response for isotopically pure graphite and graphite with a naturally abundant isotope concentration at 100 K for a 10 µm grating period at 100 K in the 1D TTG geometry.